\documentstyle[preprint,prd,aps,floats,tighten,draft,epsf]{revtex}
\newcommand{\insertplot}[1]{\epsfxsize=3.1in \epsfysize=2.6in \epsfbox{#1} }
\begin{document}
\thispagestyle{empty}
\draft
\title{Gauge-boson scattering signals at the LHC}
\author{Suraj N. Gupta and James M. Johnson}
\address{ Department of Physics, Wayne State University, Detroit,
	Michigan 48202}
\author{Glenn A. Ladinsky and Wayne W. Repko}
\address{Department of Physics and Astronomy, Michigan State University,
	East Lansing, Michigan 48824}

\date{October 3, 1995}
\maketitle
\begin{abstract}
\baselineskip=24pt

	We have extended our earlier treatment of the gauge-boson
scattering with radiative corrections in the standard model at
supercollider energies, and computed the rates for gauge-boson
scattering modes in $pp$ collisions leading to the final states
$W^+W^-$, $ZZ(4l)$, $ZZ(2l2\nu)$,
$W^\pm Z$, and $W^\pm W^\pm $.  Our results at the LHC \hbox{energy}
of $\sqrt{s}=14$~TeV for $m_H=1000$~GeV are compared with those
recently obtained by Bagger {\it et al.}  These results will be useful
in the search for the Higgs bosons at supercollider energies as well as
for experimentally distinguishing the standard model from
non-minimal Higgs models.

\end{abstract}
\pacs{13.85.Qk,14.80.Er,14.80.Gt}

\clearpage
\setlength{\parindent}{24pt}
\baselineskip=24pt
\section{INTRODUCTION}

	Along with the search for the elusive Higgs bosons, efforts
continue to be made to experimentally distinguish between the minimal
(standard) and non-minimal Higgs models\cite{gunion}.  Bagger {\it
et~al.} \cite{bagger1,bagger2} have provided an interesting approach to
deal with this issue by analyzing the gauge-boson scattering signals in
hadron supercollider experiments.  They have shown that the  relative
frequencies of the gauge-boson scattering modes leading to the final
states $W^+W^-$, $ZZ(4l)$, $ZZ(2l2\nu)$, $W^\pm Z$, and $W^\pm W^\pm $
are significantly different in the standard model from those involving
other electroweak symmetry breaking mechanisms.

	Recently, we carried out precise calculations of the $W$, $Z$,
and Higgs-boson scattering amplitudes in the standard model at
supercollider energies with the inclusion of the full one-loop
radiative corrections, and explored the detection of the Higgs boson
through gauge-boson scattering in $pp$ collisions\cite{gjr}.  We also
emphasized that the apparent violation of unitarity in scattering,
encountered by some earlier authors, is not a consequence of the fact
that the weak interactions become strong at supercollider energies.
Rather, the Feynman expansion of the scattering operator yields
non-unitary amplitudes at all energies, and the problem becomes
accentuated at higher energies.  The unitarity of our amplitudes was
ensured by using the K-matrix as well as the Pad\'e methods.

	We shall now extend our earlier treatment, and use 
unitarized scattering amplitudes to compute the rates for the various
gauge-boson scattering modes in $pp$ collisions at the LHC energy of
$\sqrt{s}=14$~TeV for $m_H=1000$~GeV.  Our results for the standard
model will be compared with those given by Bagger {\it et
al.}\cite{bagger2}

\section{Unitarized Gauge-Boson Scattering Amplitudes}

	In Ref.~4, we calculated the $S$-wave amplitudes for the
scattering of the $W$, $Z$, and Higgs-bosons in the coupled neutral
channels $W_L^+W_L^-$, $Z_LZ_L$ and $HH$.  We have now extended our
calculations by including the additional channels $W_L^\pm Z_L$ and
$W_L^\pm W_L^\pm $ as well as the $P$ waves for all the channels.  In
view of an SO(3) symmetry associated with the gauge-boson interactions,
all amplitudes are expressible in terms of three amplitudes $M$, $M'$,
and $M''$, which are given in Ref.~4.

	The $l$-wave projection of the amplitude $M(s,t,u)$ is given
by \cite{pnote}
\begin{equation}
a_l(s)=\frac{1}{32\pi}\left(\frac{4 |{\bf p}_f| |{\bf p}_i|}{s}\right)^{1/2}
        \int_{-1}^1 d\!\cos\theta \:M(s,t,u)\:P_l(\cos\theta), \label{pwform}
\end{equation}
and let $b(s)$, $\bar{b}(s)$, $c(s)$ and $d(s)$ denote similar
projections of $M(t,s,u)$, $M(u,t,s)$, $M'(s,t,u)$ and $M''(s,t,u)$,
respectively.  Then, the $l$-wave amplitude for the coupled
$W_L^+W_L^-$, $Z_LZ_L$ and $HH$ channels can be expressed as
\begin{equation}
a_l=\left( \renewcommand{\arraystretch}{1.5}
\begin{array}{ccc}
a_l+b_l&\displaystyle\frac{a_l}{\sqrt{2}}&\displaystyle\frac{c_l}{\sqrt{2}}\\
\displaystyle\frac{a_l}{\sqrt{2}}&\frac{1}{2}(a_l+b_l+\bar{b}_l)
	&\displaystyle\frac{c_l}{2}\\
\displaystyle\frac{c_l}{\sqrt{2}}&\displaystyle\frac{c_l}{2}&\displaystyle
        \frac{d_l}{2}
\end{array} \right)\ ,\label{pwamp}
\end{equation}
and
\begin{eqnarray}
a_l(W_L^\pm Z_L \rightarrow W_L^\pm Z_L)&=& b_l,\\
a_l(W_L^\pm W_L^\pm \rightarrow W_L^\pm W_L^\pm)&=&
	\frac{1}{2}(b_l+\bar{b}_l),
\end{eqnarray}
where we have introduced appropriate factors to eliminate the
identical-particle restrictions on the $W_L^\pm W_L^\pm$, $Z_LZ_L$, and
$HH$ phase spaces.

	The above amplitudes, obtained from Feynman's expansion of the
scattering operator, require unitarization, and, as explained in
Ref.~4, we have used the K-matrix and Pad\'e methods for this purpose.
The K-matrix amplitude for the $l$th partial wave is given by
\begin{equation} 
a^K_l=\frac{a^{(1)}_l+{\rm Re\,} a^{(2)}_l}{1-i\left(a^{(1)}_l+{\rm Re\,} a^{(2)}_l\right)}\ ,\label{kmatrix}
\end{equation}
while the Pad\'e amplitude is
\begin{equation}
a^P_l=\frac{{a^{(1)}_l}^2}{a^{(1)}_l-a^{(2)}_l}\ ,\label{pade}
\end{equation}
where $a^{(1)}_l$ and $a^{(2)}_l$ denote the tree and one-loop
amplitudes.  When $a_l$ is a matrix for multichannel scattering,
(\ref{kmatrix}) and (\ref{pade}) take the forms
\begin{eqnarray}
a^K_l&=&\left(a^{(1)}_l+{\rm Re\,} a^{(2)}_l\right)
        \left[1-i\left(a^{(1)}_l+{\rm Re\,} a^{(2)}_l\right)\right]^{-1},\\
a^P_l&=&a^{(1)}_l\left(a^{(1)}_l-a^{(2)}_l\right)^{-1}a^{(1)}_l.
\end{eqnarray}

	We have plotted in Fig.~1 absolute values of the K-matrix 
and the Pad\'e amplitudes for scatterings among the various gauge-boson
channels as functions of the center-of-mass energy $\sqrt{s}$ for
$m_H=1000$~GeV.  Both the $S$- and $P$-wave amplitudes are shown, and
it may be noted that the $P$-wave amplitudes vanish for states
containing identical gauge bosons.  The K-matrix and the Pad\'e
amplitudes agree at lower energies, but the Pad\'e amplitudes seem to
exhibit an interesting but odd behavior at higher energies.  In
particular,  striking $P$-wave resonances in the $W_L^+W_L^-$
and $W_L^\pm Z_L$ channels appear in
the Pad\'e amplitudes around $\sqrt{s}=3000$~GeV \cite{repko}.  The physical
significance of the Pad\'e amplitude resonances has been analyzed in
general by Atkinson, Harada, and Sanda\cite{atkinson}.

\section{Gauge-Boson Pair Production via Fusion in $\lowercase{pp}$ 
	Collisions}

	We have used the unitarized amplitudes to compute the 
invariant-mass distributions for gauge-boson pair production via fusion
in $pp$ collisions at the LHC energy of $\sqrt{s}=14$~TeV for
$m_H=1000$~GeV.  Our calculations are based on the effective-W
approximation \cite{effw}, and we have used the same cuts and
backgrounds as proposed
by Bagger {\it et al.}  The
distributions for production modes leading to the final states
$W^+W^-$, $ZZ(4l)$, $ZZ(2l2\nu)$, $W^\pm Z$, and $W^\pm W^\pm$ are
shown in Fig.~2.

	We have also computed the event rates per LHC-year for gauge-boson
fusion signals, assuming $\sqrt{s}=14$~TeV and an annual luminosity of
100~fb$^{-1}$.  Our results for the standard model, obtained with the
K-matrix and the Pad\'e unitarizations, are compared with those of
Bagger {\it et al.} in Table~I.  It is interesting that the K-matrix 
and Pad\'e unitarization results are practically the same 
for all final states despite the appearance of the $P$-wave resonances
in the channels leading to the $W^+_LW^-_L$ and $W^\pm_LZ_L$ states.

	In Ref.~3, it was shown that models for strongly interacting
symmetry breaking with a scalar isospin-zero resonance (standard model,
scalar model and O(2N) model) will yield a large excess of events in
the $Z_LZ_L\rightarrow 2l2\nu$, $Z_LZ_L\rightarrow 4l$, and
$W_L^+W_L^-\rightarrow2l2\nu$ final states, while models with a heavy
isospin-one vector resonance, or no resonance at all, will provide a
large enhancement of the $W_L^\pm W_L^\pm \rightarrow 2l2\nu$ final
states.  It is the distinguishing features of these models, examined
from the perspective of all the various final states, that allows
different mechanisms for strongly interacting electroweak symmetry
breaking to be recognized.

	Our standard-model calculations have produced some differences with
respect to the previous analysis of the electroweak symmetry breaking
sector given in Ref.~3.  The results 
computed here demonstrate an enhancement of the $W_L^\pm
W_L^\pm$ final states, bringing the event rate closer to that of the
vector models with a heavier resonance. Although this increase to about
8 events/LHC-year is still smaller than the isovector and nonresonant
enhancements computed in Ref.~3, it can still increase the expected
time it may take to distinguish these symmetry breaking mechanisms from
the standard model by a couple of years.

	Through the application of both the K-matrix and Pad\'e 
unitarizations, we find some decrease in the event rates for the
$W_L^+W_L^-$, $Z_LZ_L(4l)$, and $Z_LZ_L(2l2\nu)$ modes. This does
not significantly affect
the usefulness of these modes in distinguishing the standard model or O(2N) models
from the other mechanisms of spontaneous symmetry breaking; there is
only an indication that the separation between the chiral lagrangian
model with a heavy scalar resonance and the standard model becomes less
clear.

	As for the production of $W_L^\pm Z_L$ pairs, with the K-matrix 
or the Pad\'e unitarization there is no significant change
from the results of Ref.~3.

	Since the standard model is the most popular and economical of all
the models, it is desirable to obtain the standard model results as
precisely as possible.
Our theoretical results will be useful in the search for the Higgs
bosons at supercollider energies as well as for experimentally
distinguishing the standard model from non-minimal Higgs models.

\acknowledgments
	This work was supported in part by the U.S. Department of Energy
under Grant No.~DE-FG02-85ER40209 and the National Science Foundation
under Grant Nos.~PHY-93-07980 and PHY-93-19216.

\baselineskip=12pt
\setlength{\unitlength}{1.in}
\begin{figure}[p]
\begin{picture}(6.4,8.0)
\put(0.,5.5){\insertplot{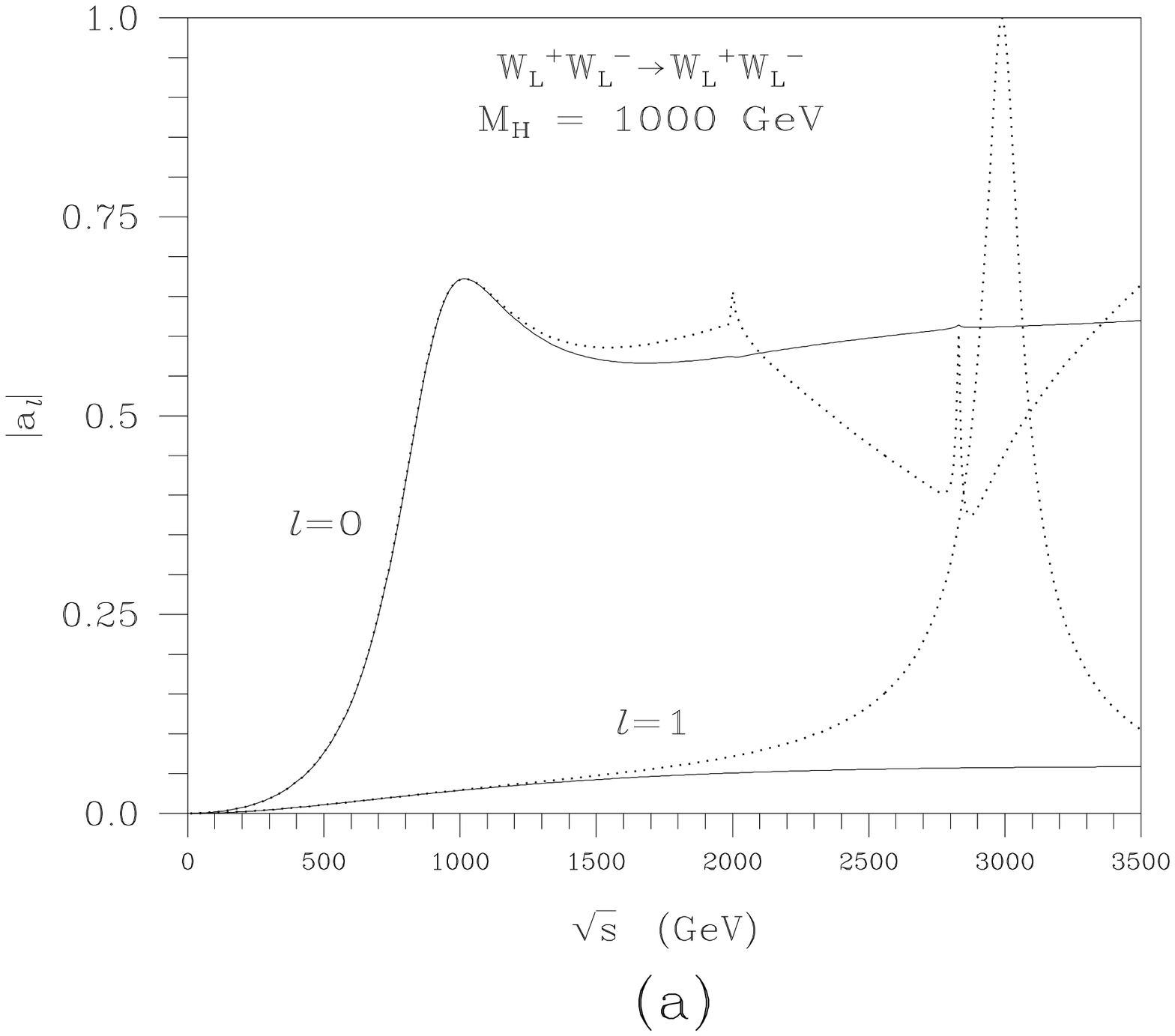} }
\put(3.5,5.5){\insertplot{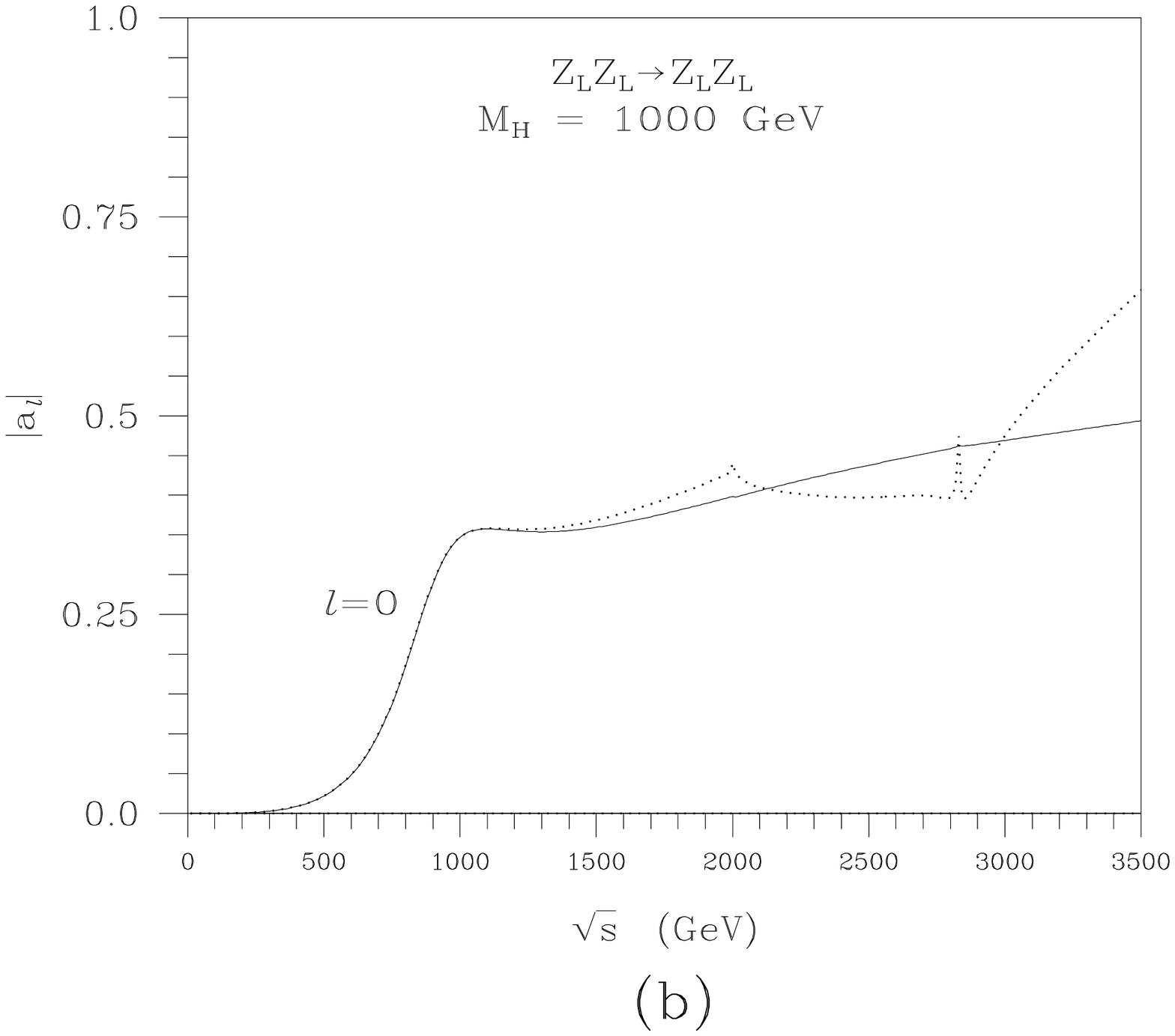} }
\put(0.,2.8){\insertplot{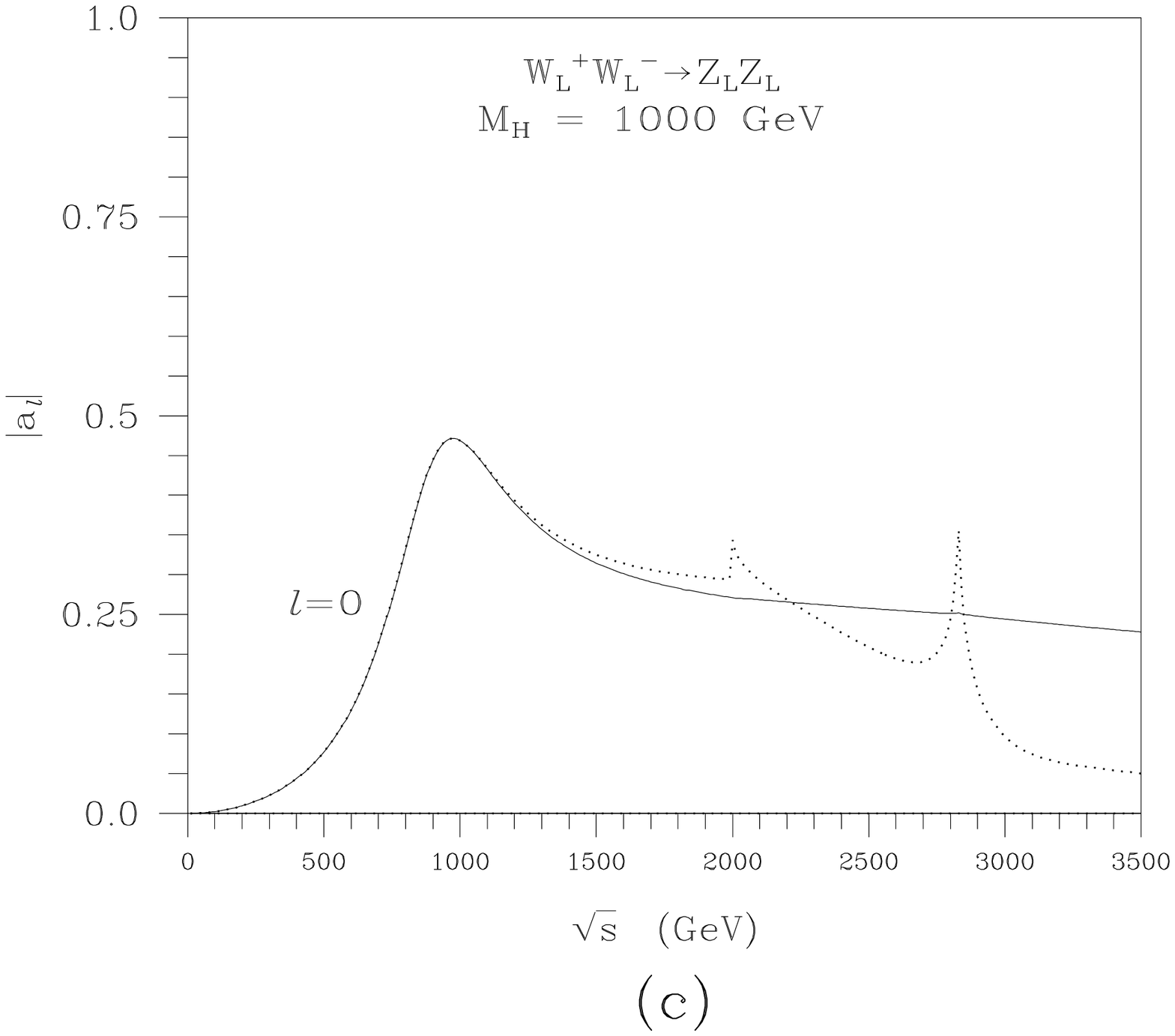} }
\put(3.5,2.8){\insertplot{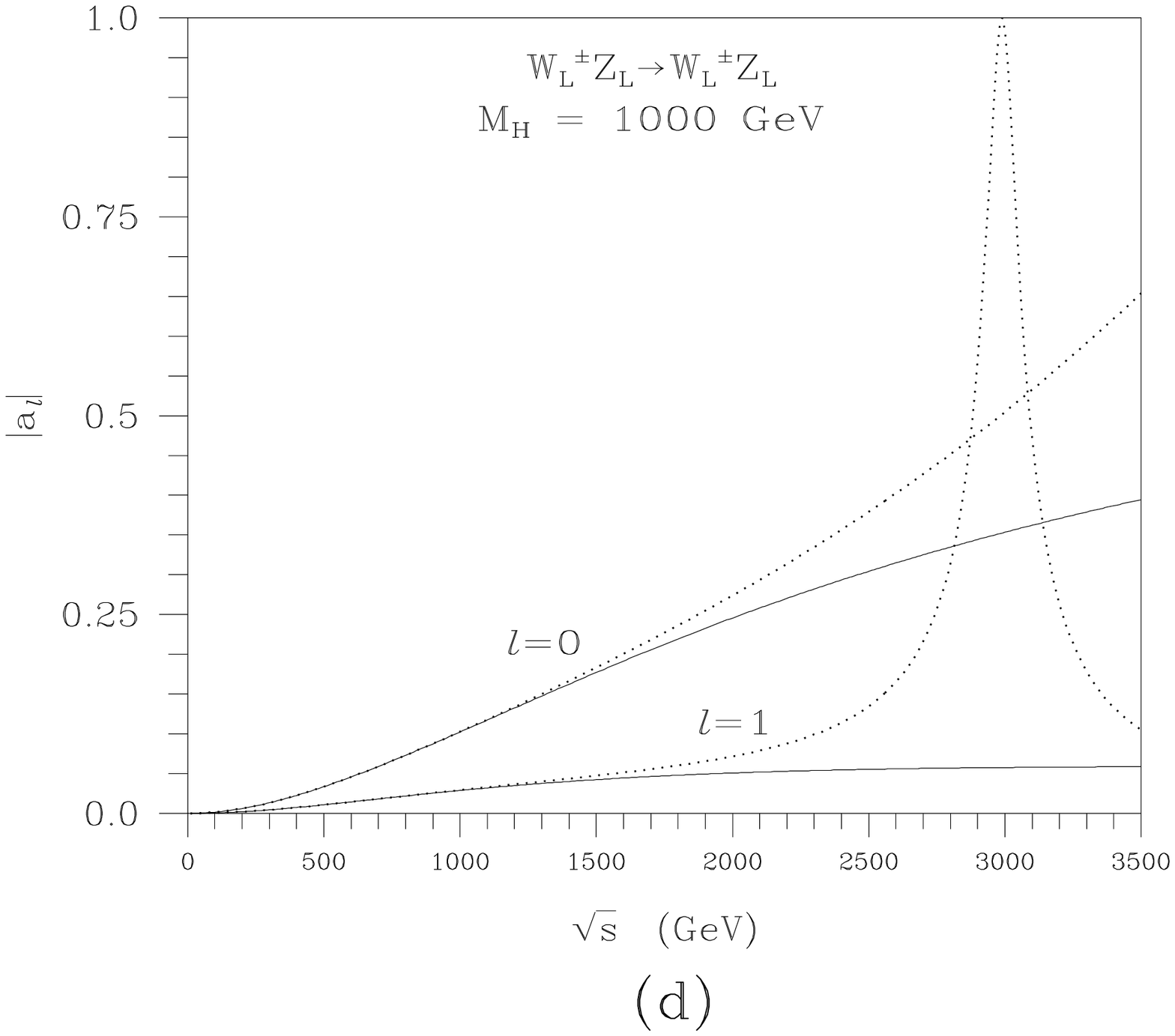} }
\put(1.75,0.15){\insertplot{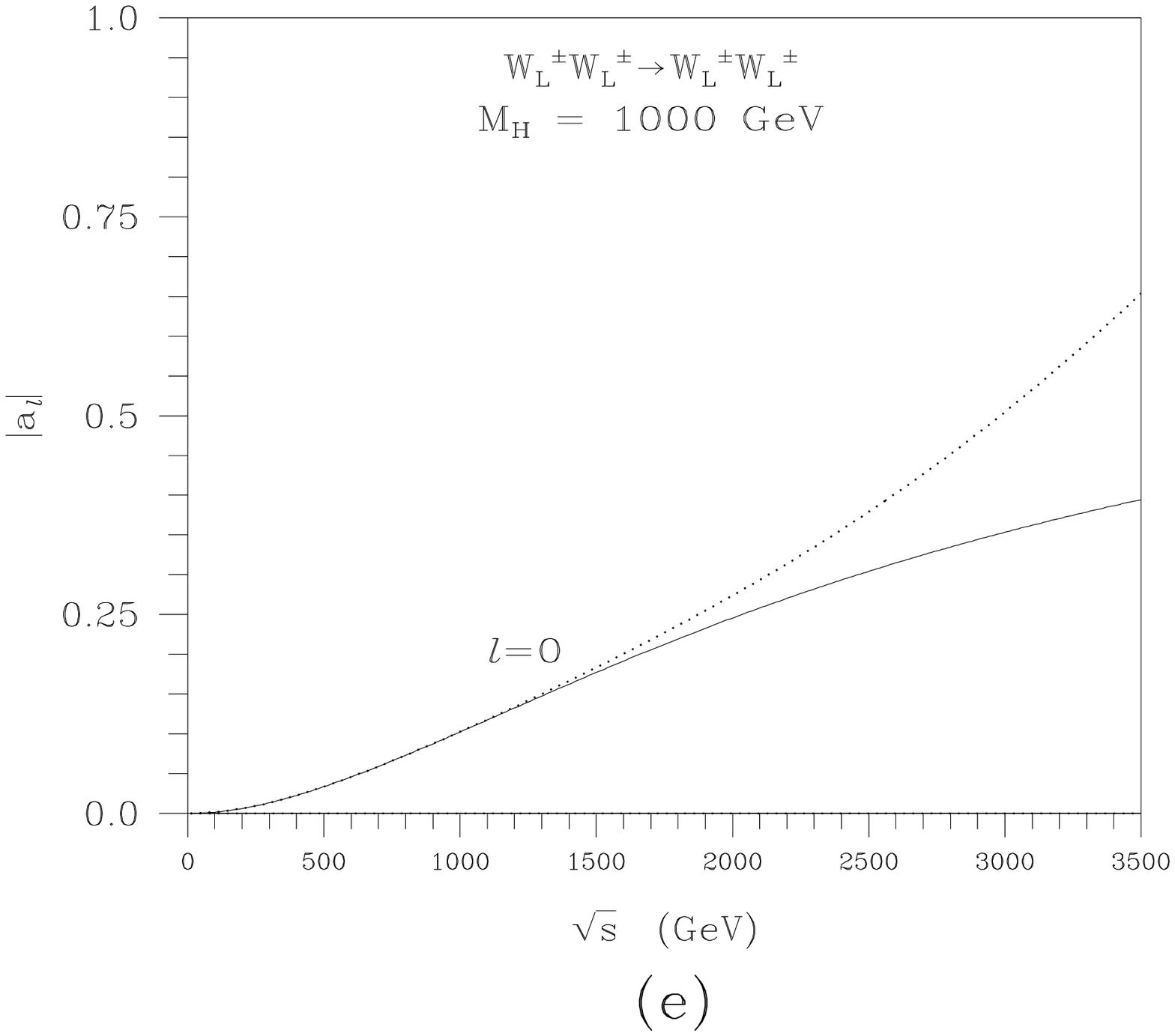} }
\end{picture}
\caption{ Absolute values of the $S$- and $P$-wave amplitudes for scattering
among the gauge-boson channels as functions of the center-of-mass energy
$\protect\sqrt{s}$ for $m_H=$1000~GeV. The solid lines represent the 
K-matrix amplitudes, and the dotted lines the Pad\'e amplitudes. 
Note that $P$-wave amplitudes vanish for states containing identical
gauge-bosons.}
\end{figure}
\begin{figure}[p]
\begin{picture}(6.4,7.5)
\put(0.,5.3){\insertplot{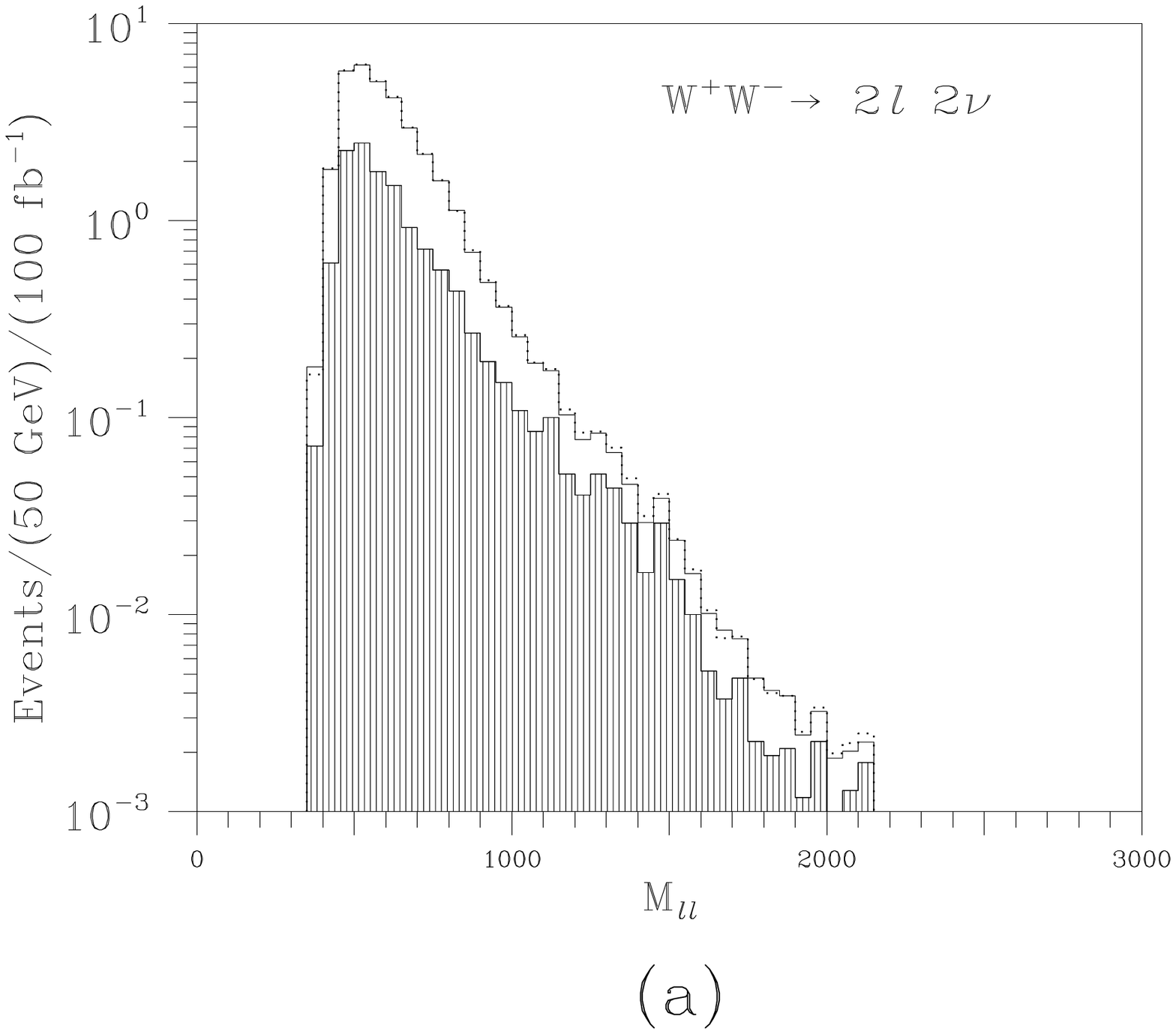} }
\put(3.5,5.3){\insertplot{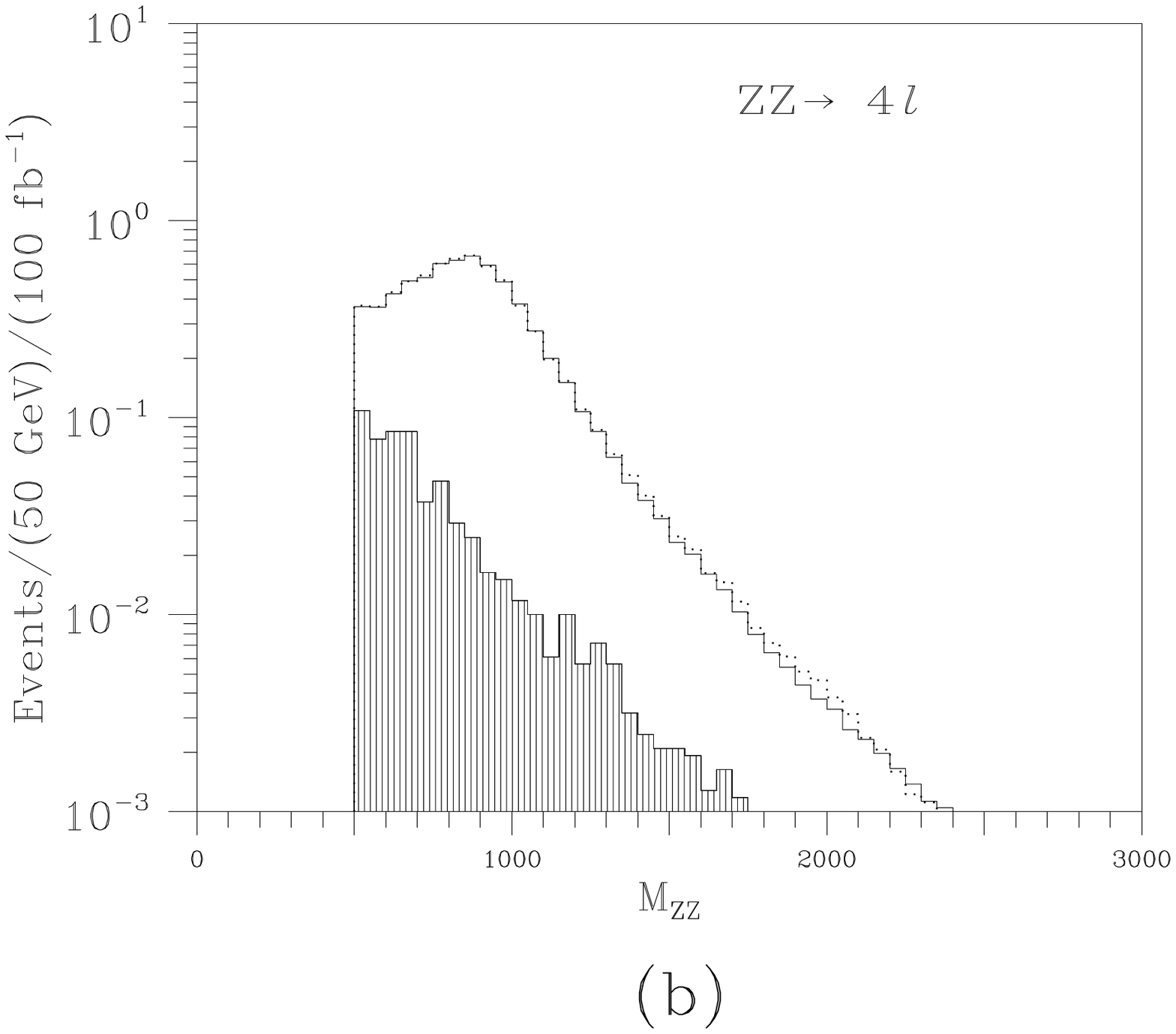} }
\put(0.,2.7){\insertplot{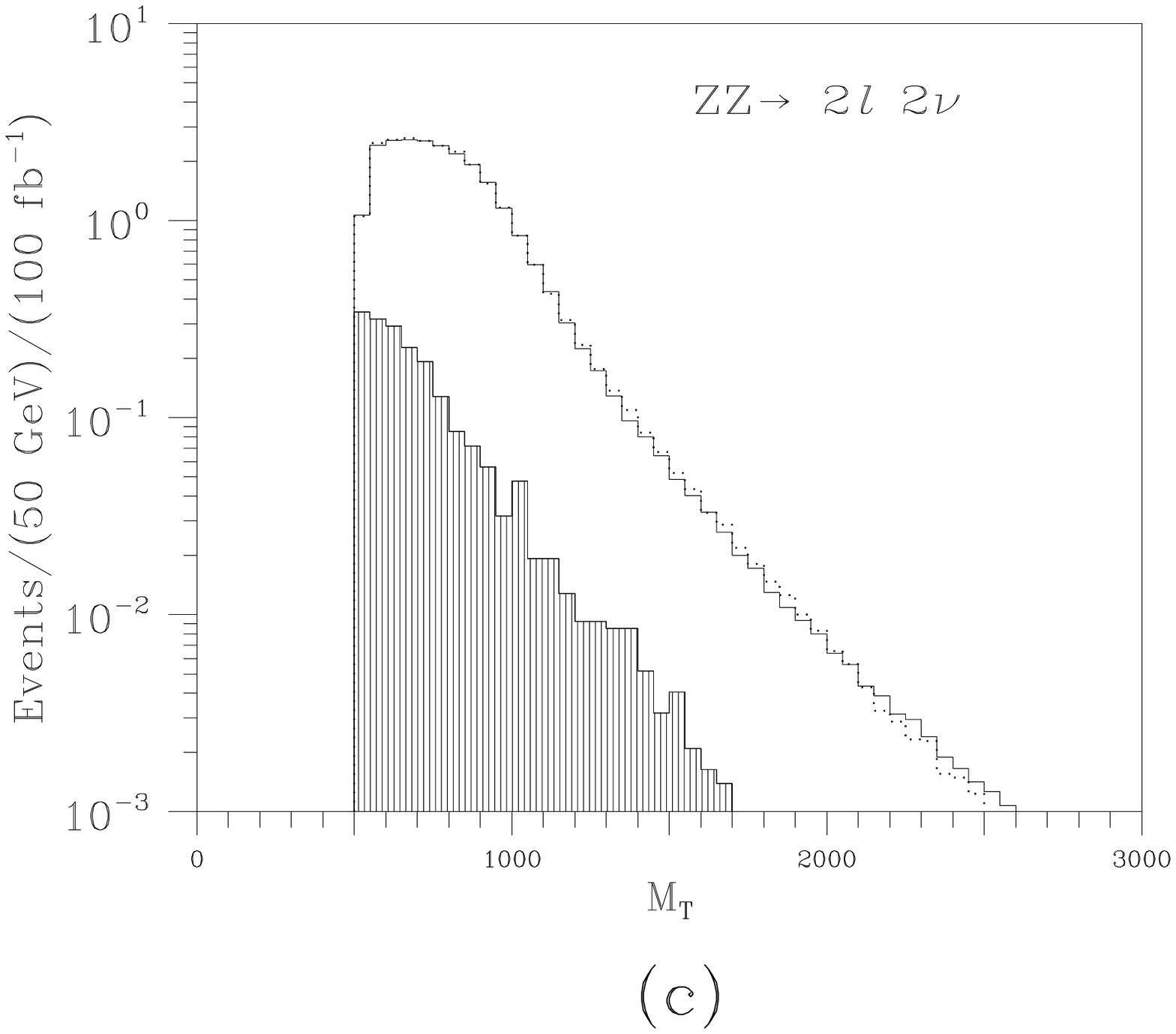} }
\put(3.5,2.7){\insertplot{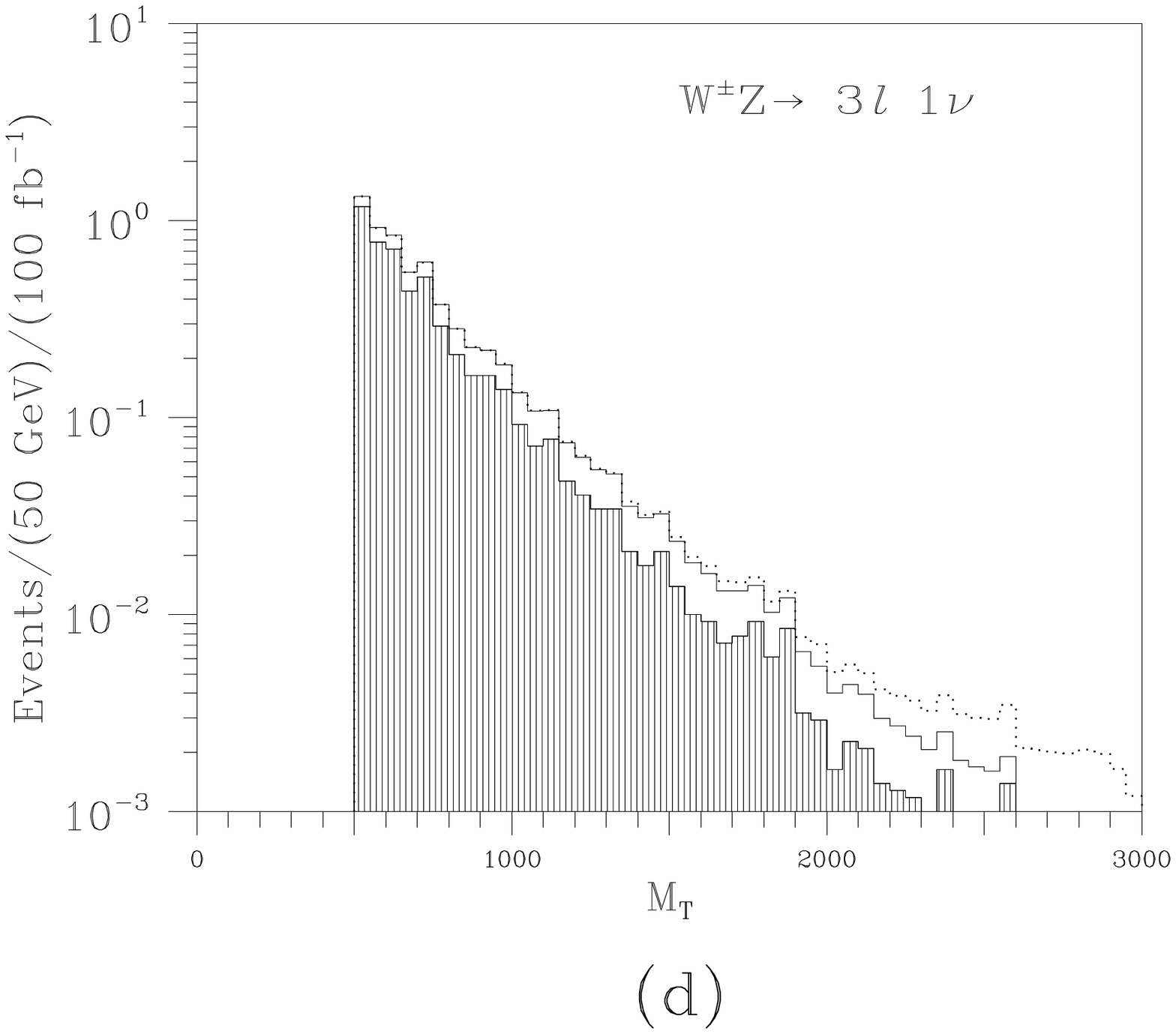} }
\put(1.8,0.1){\insertplot{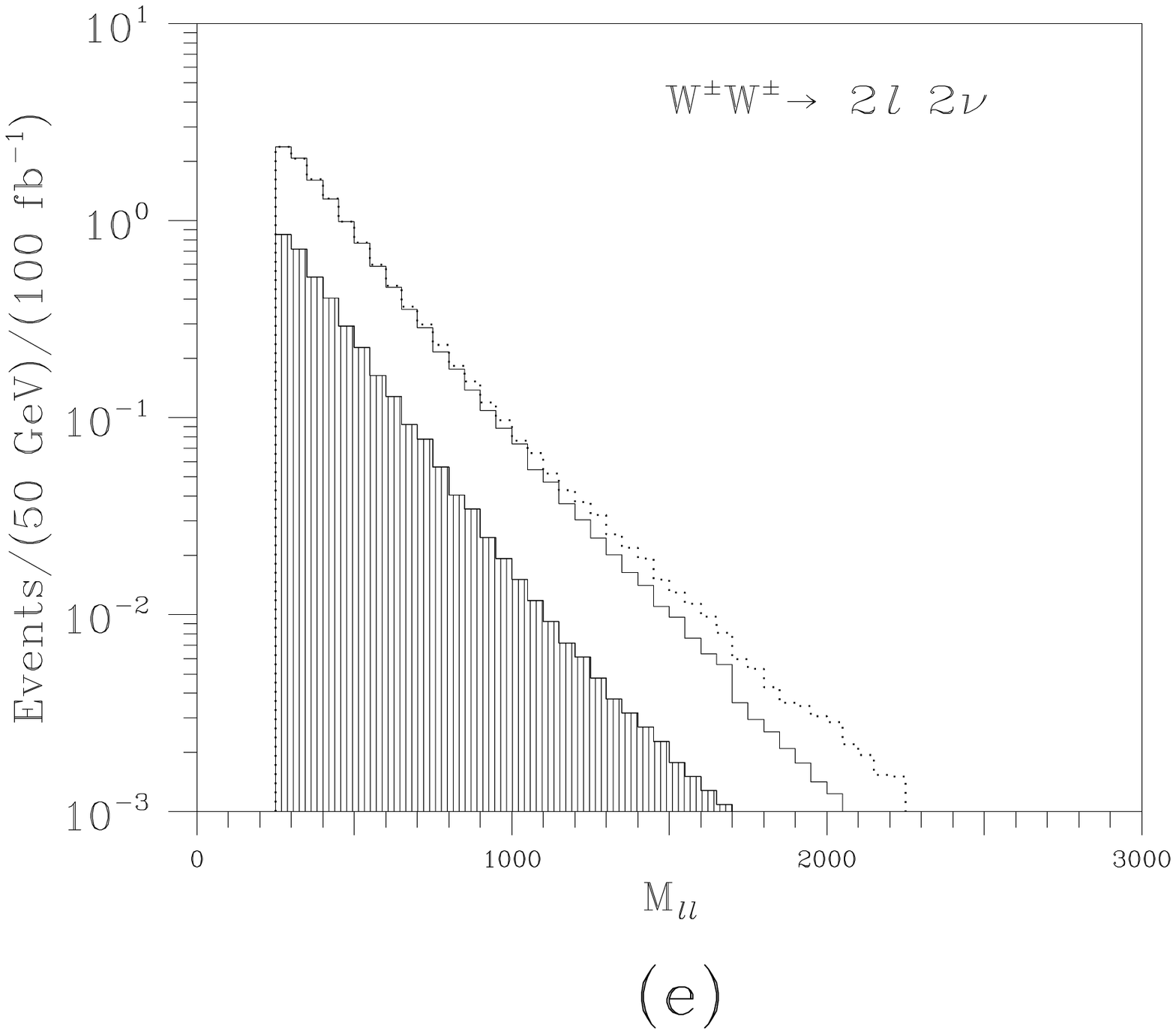} }
\end{picture}
\caption{ Invariant mass distributions for the purely leptonic final
states arising from the processes $pp\rightarrow W^+W^-X\rightarrow
2l2\nu X$, $pp\rightarrow ZZX\rightarrow 4lX$, $pp\rightarrow
ZZX\rightarrow 2l2\nu X$, $pp\rightarrow W^\pm ZX\rightarrow 3l\nu X$
and $pp\rightarrow W^\pm W^\pm X\rightarrow 2l2\nu X$ for
$\protect\sqrt{s}=14$~GeV and an annual LHC luminosity of
100~fb${}^{-1}$. The K-matrix and Pad\'e signals for the standard model
with $m_H=1000$~GeV are shown as solid and dotted lines above the
background, and plotted as the number of events per 50~GeV bin, per
100~fb${}^{-1}$.  $M_{ll}$ is the invariant mass of the experimentally
observed leptons, $M_{ZZ}$ is the invariant mass of the $ZZ$ pair, and
$M_T$ is the cluster transverse mass as defined in Ref.~3. }
\end{figure}
%
\begin{table}
\caption{ Event rates per LHC-year for gauge-boson fusion, together 
with backgrounds, assuming $\protect\sqrt{s}=14$~TeV and an annual luminosity
of 100~fb$^{-1}$.  Our cuts and backgrounds are the same as those of
Bagger {\it et al.} }
\bigskip
\begin{tabular}{ldddd}
Final state&\multicolumn{1}{c}{Bkgd.}&\multicolumn{1}{c}{Bagger {\it et. al}}&	
	K-Matrix&	Pad\'e\\
	&		&		&	unitarization&	unitarization\\
\tableline
$W^+W^-$&	12&	27&	21& 	21 \\
$ZZ(4l)$&	0.7&	9&	6.1& 	6.1 \\
$ZZ(2l2\nu)$&	1.8&	29&	21& 	22 \\
$W^\pm Z$&	4.9&	1.2&	1.3& 	1.3 \\
$W^\pm W^\pm$&	3.7&	5.6&	8.2& 	8.4 \\
\end{tabular} 
\end{table}

\begin{thebibliography}{99}
\bibitem{gjr} S. N. Gupta, J. M. Johnson and W. W. Repko, Phys. Rev. D
	{\bf 48}, 2083 (1993).  Extensive references to earlier work
	are cited in this paper.  Note that in
	the formula for the integral $B_3(s)$, given in the Appendix of
	the paper, ``$2-\frac{2\pi}{\sqrt 3}$ for $s=1$'' should read
	``$2-\frac{\pi}{\sqrt 3}$ for $s=1$''.
\end{thebibliography}
\end{document}